# Secure Communication Scheme Based on Asymptotic Model of Deterministic Randomness

Jiantao Zhou *, Wenjiang Pei, Kai Wang, Jie Huang, Zhenya He

*Department of Radio Engineering, Southeast University, Nanjing, 210096, China*

## Abstract

We propose a new cryptosystem by combing the Lissajous map, which is the asymptotic model of deterministic randomness, with the one-way coupled map lattice (OCML) system. The key space, the encryption efficiency, and the security are investigated. We find that the parameter sensitivity can reach the computational precision when the system size is only three, all the lattice outputs can be treated as key stream parallelly, and the system is resistible against various attacks including the differential-like chosen cipher attack. The findings of this paper are a strong indication of the importance of deterministic randomness in secure communications.

## 1. Introduction

Due to the intrinsic connection to cryptography, chaos represents a good candidate for designing cryptosystems, especially since the first description of chaos synchronization by Pecora and Carroll [1]. A large number of chaos-based secure communication schemes have been proposed in last decade [2-8]. However, most of them are fundamentally flowed by a lack of security, and various attacks such as nonlinear forecasting, return map, adaptive parameter estimation, error function attack (EFA), and inverse computation based chosen cipher attack, can succeed in extracting the secret keys or recovering the messages encoded by the chaotic waves directly [9-13]. Recently, a class of spatiotemporally chaotic systems with high security and efficiency by applying some algebraic operations into the one-way coupled map lattice (OCML) system has been provided [5-8]. To date the exact mechanism for high security, however, has not been revealed yet. In this study, we demonstrate their outstanding features, excellent chaos [14], presents fundamentally differences with known chaotic dynamics—directly linked with deterministic randomness firstly introduced in Ref. [15], sharing short time unpredictability, longer period, and better random properties. In other words, the internal states are transformed into a very complex dynamics, i.e., key streams, by non-linear functions with non-invertible characteristics such as INT and MOD operations [5-8]. However, we have found that this class of cryptosystem still suffers from the problem of degeneration, and the differential-like chosen cipher attack (DCCA) can break the prototype of such cryptosystems with ignorable computational costs [16]. With the help of our recently proposed Lissajous map: the exact model of asymptotic deterministic randomness constructed by the skewed parabola map and non-invertible nonlinear transformation [17-18], the current study examines the alternative scenario of cryptosystem under such novel nonlinear dynamics in a systematic way. A one-way coupled Lissajous map lattice (OCLML) with a delicately designed confusion unit in every lattice is implemented to achieve significant enhancement in confusion and diffusion. The key space, the encryption efficiency, and the security are investigated in detail. Both the analytical and experimental results show that the parameter sensitivity can reach the level of $2^{-52}$, the computational precision in practice, when the system size is only three. The parallel encryption can be realized even in the one-dimensional coupling structure. Furthermore, the cryptosystem can resist almost all the existing attacks including EFA, DCCA etc.

* Corresponding author. E-mail address: jtzhou@seu.edu.cn.

This paper is organized as follows. In section 2, we briefly introduce deterministic randomness. In section 3, we present the OCLML system and investigate its key space and encryption efficiency. Security analysis and numerical results are shown in section 4, and some concluding remarks are given in section 5.

## 2. Deterministic randomness and its asymptotic model

Here, we provide a brief description of deterministic randomness and its asymptotic models (For more details see Ref. [15, 17-20] and references therein). Originally, Gonzalez found that the generalization of the solution to the well known logistic map $X_n = \sin^2(\theta\pi z^n)$, where $z$ is a real number, can produce short-term unpredictable sequences [15, 19-20]. In order to distinguish from chaos, this phenomenon is named as deterministic randomness [20]. The basic concept of deterministic randomness is that the next value of the generated sequence cannot be expressed as $X_{n+1} = f(X_n, X_{n-1}, \cdots, X_{n-r+1})$, i.e., it is impossible to derive the next value by the previous values. Several autonomous systems and real physical systems have been provided to produce *similar* dynamics of asymptotic deterministic randomness [20]. To the best of our knowledge, Lissajous map is the first model to *exactly* describe asymptotic deterministic randomness [17-18].

**Theorem1** Consider $x_{n+1} = f(af^{-1}(x_n))$ , $a = p/q$ , $y_{n+1} = f(bf^{-1}(x_{n+1}))$ , when $b = q^N$ and $b = 2(pq)^N$ for $f := \sin^2$ , $f := \cos^2$ , $f := \cos$ and $f := \sin$ , respectively. The map of $y_n$ vs. $y_{n+i}$ , $i = 1, 2, \cdots N$ , will be bi-multivalued and represent Lissajous curves, thus name it as Lissajous map [17-18]. Furthermore, the sequences produced by Lissajous map are $m$-steps unpredictable, where $1 < m \le N$ , i.e., the next value cannot be expressed by $y_n = f(y_{n-1}, y_{n-2}, \cdots, y_{n-m})$ .

Taking the example of $f := \sin^2$, we can have $y_{n+M} = \sin^2(a^M q^N \sin^{-1}(\sqrt{x_n}))$, $M < N$, which will be equivalent to $X_{n+M} = \sin^2(\theta\pi a^{n+M})$ under certain initial conditions. Obviously, when $b \to \infty$, Lissajous map can be used to describe deterministic randomness and act as an asymptotic model [17-18]. Fig.1 shows the phase space of sequence generated by Lissajous map, which are identical to that of sequence produced by $X_n = \sin^2(\pi\theta z^n)$. The route from chaos to deterministic randomness can be visually described by the bifurcation process of Lissajous map, shown in Fig. 2, where $x_{n+1} = u\sin(a\sin^{-1}(x_n))$, $y_{n+1} = \sin(b\sin^{-1}(x_{n+1}))$ with $a = 5/2$ and $u \in [0,1]$. It can be found that the bifurcation process of Lissajous map can be empirically explained as some "frequency modulation transformation" for that of typical chaotic system, and the "instant frequency" is $b\sin(\frac{p}{q}\sin^{-1}(x_n))/\sqrt{1-(u\sin(\frac{p}{q}\sin^{-1}(x_n)))^2}$.

It should be noted that the spatiotemporally chaotic systems proposed in Ref. [5-8] also belong to special cases of deterministic randomness generator, in which the output, even some hidden state variables, is derived from the internal states by some non-invertible nonlinear functions. Meanwhile, it is interesting to find that the deterministic randomness generator is directly associated with the counter-assisted generator with provable high security in conventional cryptography [21].

## 3. Cryptosystem incorporating Lissajous map with OCML

OCML as a theoretical model of spatiotemporal phenomena has attracted much attention in secure communications [5-8]. In comparison with low-dimensional chaos, spatiotemporal chaos is much more complex in dynamics. Moreover, there exist many spatial sites, each of which taking chaotic motion can be performed parallelly to increase the efficiency of information treatment. Therefore, OCML system provides an ideal platform to be equipped with Lissajous map to build a cryptosystem with high security and high efficiency. Before presenting our scheme, let us analyze the OCML system in term of its Lyapunov spectrum (LS). Explicitly, we consider the following OCML of length $N$ with period boundary condition.

$$x_j(n+1) = (1-\varepsilon) f[x_j(n)] + \varepsilon f[x_{j-1}(n)], \quad j = 1, 2, \cdots N$$

$$x_j(n) = x_{j+N}(n) \tag{1}$$

In order to calculate the LS, we differentiate the state equation to obtain the evolution equations for tangent vectors $\zeta = (\delta x_1, \delta x_2 \ldots \delta x_N)^T$, which in matrix form read $\zeta_{n+1} = \mathbf{T}_n \zeta_n$, with the Jacobian matrix given by $\mathbf{T_n} = [(1-\varepsilon) + \varepsilon \mathbf{B}]\mathbf{D_n}$, where $D_n^{jk} = F'(x_j(n))\delta_{j,k}$, $B^{jk} = \delta_{j,(k+1)\bmod N}$.

An important case that can be tacked easily is the one where the lattices are in homogeneous state, i.e., $x_1(n) = x_2(n) = ... = x_N(n) = x^*(n)$. It can be found that this state can be obtained if we choose the initial state as $x_0(1) = x_1(1) = x_2(1) = ... = x_N(1)$. In other words, the homogeneity of the initial conditions is preserved under iterations.

Due to the homogeneity, we have $\mathbf{T_n} = [(1-\varepsilon) + \varepsilon \mathbf{B}]\mathbf{D_n} = F'(x^*(n))\hat{\mathbf{B}}$, where $\hat{\mathbf{B}} = (1-\varepsilon)\mathbf{I_N} + \varepsilon \mathbf{B}$. If $\Lambda_1, \cdots \Lambda_N$ are the eigenvalues of $\hat{\mathbf{\Lambda}} = \lim_{n\to\infty} (\mathbf{\Pi_n}\mathbf{\Pi_n^T})^{1/2n}$, where $\mathbf{\Pi_n} = \mathbf{T_n}\mathbf{T_{n-1}} \cdots \mathbf{T_2}\mathbf{T_1}$, the LS are obtained as $\lambda_k = \ln \Lambda_k$, $k = 1, \cdots, N$. Therefore, we have the LS as $\lambda_k^* = \lambda_u + \ln\left|(1-\varepsilon) + \varepsilon b_k\right|$, where $\lambda_u = <\ln\left|f'(x^*(n))\right|>$ is the Lyapunov exponent (LE) of uncoupled map, and $b_k$ is the $kth$ eigenvalue of $\mathbf{B}$. We notice that $\mathbf{B}$ is a circulate matrix, its eigenvalues are given by: $c_0 + c_1 r_j + \cdots c_{N-1} r_j^{N-1}$, where $r_j = \exp(2\pi i j / N)$ is an $N$th root of unity, and $c_j = \begin{cases} 1 & j = N-1 \\ 0 & otherwise \end{cases}$. Thus, we can derive the LS of the OCML system as:

$$\lambda_k^* = \lambda_u + \ln\left(\sqrt{(1-\varepsilon)^2 + \varepsilon^2 + 2\varepsilon(1-\varepsilon)\cos\frac{2\pi k}{N}}\right) \tag{2}$$

The largest LE (LLE) $\lambda_{\max} = \lambda_u$, which determines the expand rate with the evolution of time, is only governed by the uncoupled nonlinear function $f$. Increasing the coupling size $N$ has no contribution to enhance the LLE, i.e., only the number of positive LE increases with *N*, instead of the value of the LLE. Therefore, in order to achieve more complex dynamics, the nonlinear function $f$ with better cryptographic properties should be involved in the OCML system.

In contrast to previous studies based on chaotic systems, we introduce the deterministic randomness to the synchronization of OCML. Our aim is to design a new cryptosystem that is “truly” secure and most efficient. By this we mean cryptosystem with the property that, when evaluating with EFA, the size of key basin is of the order of the computational precision, and every state variable can be used to encrypt the message simultaneously. The dynamics of the transmitters is shown in Eq. 3. In the new OCML-based cryptosystem, every lattice consists of two parts: a coupling unit and a nonlinear transformation unit. The output of every lattice is derived from the coupling variables by a non-invertible nonlinear function. It should be noted that the output of every lattice can be used to produce key stream for encoding the messages simultaneously, which significantly increases the efficiency of encryption.

$$x_{1,1}(n+1)=(1-\varepsilon)F_1(x_{1,1}(n))+\varepsilon F_1(x_{0,1}(n)) \tag{3a}$$

$$x_{1,2}(n+1)=G_1(x_{1,1}(n+1)) \tag{3b}$$

$$x_{j,1}(n+1)=(1-\varepsilon)F_j(x_{j,1}(n))+\varepsilon F_j(x_{j-1,2}(n)) \tag{3c}$$

$$x_{j,2}(n+1)=G_j(x_{j,1}(n+1))\text{，}\quad j=2,...,N \tag{3d}$$

$$K_n(j)=[\text{int}(x_{j,2}(n)\times 2^{\mu})]\bmod 2^{\gamma} \tag{3e}$$

$$x_{0,1}(n)=S_n/2^{\gamma} \tag{3f}$$

with $F_j(x)=f(a_j f^{-1}(x))$，$G_j(x)=f(b_j f^{-1}(x))$，$f(x)=\sin^2(x)$.

The dynamics of the receiver (denoted by $y_{i,j}(n)$, $i=1,2,...,N$, $j=1,2$) is identical to that of the transmitter except that the first lattice $y_{1,1}(n)$ is driven by $y_{0,1}(n)=x_{0,1}(n)$.

In our model we fix $\varepsilon = 0.99$, $b_j = 2^{52}$, $j = 1,2,...,N$, $N = 3$, $\mu = 52$, $\gamma = 32$, $a_j = 2.5, j = 2,3,...,N$, and adopt $a_1$ as the secret key.

The transmitted ciphertext has two functions: carry the information of plaintext and drive the decrypter to synchronize. When they are synchronized, the decrypter recovers the key stream as that generated in the encrypter. From the state equation, we can derive the Jacobian matrix for the driven sub-system, which is a trigonometric matrix with trigonometric variables:

$$(D\mathbf{F})_{ii} = (1-\varepsilon)F_i^{'}(y_i),\quad i = 1,2,...,N \tag{4}$$

As a result, the conditional LE read:

$$\begin{aligned}\lambda_i &= \ln(1-\varepsilon) + \lim_{T\to\infty}\frac{1}{T}\sum_{t=1}^{T}\ln(F_i^{'}(y_i(t))) \\ &= \ln(1-\varepsilon) + \lim_{T\to\infty}\frac{1}{T}\sum_{t=1}^{T}\ln\left|\frac{a_i\sin(2a_i\arcsin\sqrt{y_i(t)})}{2\sqrt{y_i(t)(1-y_i(t))}}\right|\end{aligned} \tag{5}$$

Only when all the conditional LEs are negative, the two sides can achieve stable synchronization. From Fig. 3 we can see that when the coupling strength is strong, the secret key can be selected in a wide range, making the whole system synchronize. Therefore, in our model, the key space is defined as $a_1 \in [2.5,100]$. But for rapid as well as stable synchronization, the selection of $a_1$ should make the largest conditional LE far away for the zero point.

A crucial condition for the multi-channel communications is that any two outputs serving as the key stream from different transmission channel should be unrelated each other. Given two sequences $H^i(N)$ and $H^j(N)$, the normalized auto-correlation and cross-correlation are, respectively, defined as follows:

$$\begin{aligned}C_{ii}(\tau) &= \frac{\sum_{l=1}^{T}[H^i(l)-\bar{H}^i][H^i(l+\tau)-\bar{H}^i]}{\sum_{l=1}^{T}[H^i(l)-\bar{H}^i]^2} \\ C_{ij}(\tau) &= \frac{\sum_{l=1}^{T}[H^i(l)-\bar{H}^i][H^j(l+\tau)-\bar{H}^j]}{\sum_{l=1}^{T}[H^i(l)-\bar{H}^i]^2}\end{aligned} \tag{6}$$

where $\bar{H}$ is the average of $H(l)$ in $T$ iterations.

In Fig. 4a, we plot the cross-correlation of the key stream produced by two adjacent sites of the cryptosystem proposed in Ref. [5]. It can be found that there exists strong correlation between key streams, which cannot be simultaneously treated to encode the messages. In comparison, we plot in Fig. 4a the cross-correlation of the key stream generated by our scheme, from which we can see that the key stream generated from different sites is practically unrelated in the sense that their cross-correlation value is equivalent to that of two sequences uniformly distributed in $[0,1]$. This property of cross-correlation is quite desirable to ensure security and overcome the interferences in a multi-channel communication environment.

## 4. Security analysis and numerical simulation

With the requirement of public-structure and chosen-cipher, EFA can be used for evaluating the system’s key sensitivity [12].

$$e(b_1) = \frac{1}{2^v T} \sum_{n=1}^{T} \left| I'_n - I_n \right| \tag{7}$$

where $I'_n$ can be computed by the intruder from the receiver with designed ciphertext $S_n$ and the test key $b_1$. Since the test key, which is close enough to the true secret key, can still synchronize the receiver to a certain extent, thus forming a key basin around the secret key.

In Fig.5a, we plot the EFA result of the model suggested in Ref. [5] with respect to the test key mismatch. We find that the width of the smooth basin around the secret key is to the level of $10^{-7}$. Once the location of the key basin is found, the attack can implement an optimal searching method to derive the actual secret key. As a comparison, we also plot in Fig.5b the EFA result of our system. It can be found that the width of the key basin is of the same order of the computational precision $10^{-16}$. Moreover, away from the basin, the error function $e(b) \approx 1/3$, practically similar to the average error between two completely random data sequences which are uniformly distributed in [0,1], just with a small fluctuation. So, it is impossible for the intruder to find the tendency toward the position of the key by certain optimal searching methods. Meanwhile, we plot the plaintext recovery in Fig. 6, from which we can see that even when there is a mismatch in the secret key of $2^{-51}$, the plaintext recovery will fail. Therefore, the proposed scheme possesses much better security against the EFA, i.e., the key sensitivity has been increased significantly.

In Ref. [16], we introduced differential mechanism into the constant-driving chosen-cipher attack, and suggested the DCCA, which can break the cryptosystem proposed in Ref. [5]. By using DCCA, we use a constant driving $y_d$ to make the chaotic system converge, and after convergence, we perturb the constant driving slightly. If we take the key stream before and after the perturbation as $K$ and $K^{'}$ respectively, we can construct a perturbation function $g(b, y_d) = K^{'} - K$, which represents the difference before and after the convergence in term of the secret key and the driving. The vulnerability of the system proposed in Ref. [5] is due to the fact that when the driving is properly selected, the perturbation function is simple structured and slowly changing [16]. Thus, some optimal searching methods based on the perturbation function can be implemented to search the actual secret key. Now we analyze the security of our proposed system under such DCCA. The key point is to analyze the complexity of the perturbation function, which can be characterized by the number of zero-crossing points. We take $F_1(x_{0,1}(n))$ as the equivalent driving, i.e., $x_d = F_1(x_{0,1}(n))$. It can be easily proved that driven by a constant signal, all the state variables will converge to certain fix points, i.e., all the state variables are independent with time index after convergence. Therefore, the state equations after the perturbation can be shown as follows.

$$x'_{1,1} = (1-\varepsilon)F_1(x_{1,1}) + \varepsilon(x_d + \tau) \quad (8a)$$

$$x'_{1,2} = G_1(x'_{1,1}) \quad (8b)$$

$$x'_{j,1} = (1-\varepsilon)F_j(x_{j,1}) + \varepsilon F_j(x'_{j-1,2}) \quad (8c)$$

$$x'_{j,2} = G_j(x'_{j,1}),\ j = 2...N \quad (8d)$$

Therefore, the difference dynamics becomes:

$$e_{1,1} = x'_{1,1} - x_{1,1} = \varepsilon\tau,\ \ e_{1,2} = x'_{1,2} - x_{1,2} = G_1(x'_{1,1}) - G_1(x_{1,1}) \quad (9a)$$

$$e_{j,1} = \varepsilon[F_j(x'_{j-1,2}) - F_j(x_{j-1,2})],\ \ e_{j,2} = G_j(x'_{j,1}) - G_j(x_{j,1}) \quad (9b)$$

As the perturbation is extremely slight, $(x'_{i,j} - x_{i,j})^2 = 0$ in practical calculation, we have

$$e_{1,1} = \varepsilon\tau,\ \ e_{1,2} = G'_1(\xi_1)e_{1,1} \quad (10a)$$

$$e_{j,1} = \varepsilon F'_j(\varsigma_j)e_{j-1,2},\ \ e_{j,2} = G'_j(\xi_j)e_{j,1},\ \ j = 2,3,\cdots,N \quad (10b)$$

where $\varsigma_j \in (\min(x'_{j,1}, x_{j,1}), \max(x'_{j,1}, x_{j,1}))$

$\xi_j \in (\min(x'_{j,2}, x_{j,2}), \max(x'_{j,2}, x_{j,2}))$, $j = 1,2,\cdots,N$

Consequently, we obtain the perturbation function

$$g(b, y_d) = 2^{\mu} e_{N,2} = \left(\prod_{j=1}^{N} G'_j(\xi_j) \prod_{j=2}^{N} F'_j(\varsigma_j)\right)\varepsilon^N 2^{\mu}\tau \quad (11)$$

Since $F'_j(x) = \frac{a_j \sin(2a_j \arcsin\sqrt{x})}{2\sqrt{x(1-x)}}$, $G'_j(x) = \frac{b_j \sin(2b_j \arcsin\sqrt{x})}{2\sqrt{x(1-x)}}$, which, respectively, has $\lfloor a_j \rfloor + 1$ and $\lfloor b_j \rfloor + 1$ zero-crossing points. We can roughly calculate the number of zero-crossing points of the perturbation function as $[(\lfloor b_j \rfloor + 1)(\lfloor a_j \rfloor + 1)]^N \approx 2^{54N}$. We can see that even when $N = 1$, the number of zero-crossing points is extremely large. Therefore, by using the perturbation function, the attacker still find it extremely difficult to construct an optimal searching method to search the secret key. In comparison with the results in Ref. [16], we plot the in Fig.7 the converged state with driving around 0.495. We see that even in the interval with $10^{-10}$ width, the function is extremely rapid changing. It indicates that the structure of the perturbation function is quite complex. Therefore, the resistance of our proposed scheme to the DCCA has been increased dramatically.

## 5. Conclusion

In summary, we investigate the deterministic randomness and its asymptotic model Lissajous map. A new cryptosystem incorporating Lissajous map with OCML has been proposed, which combines the features of deterministic randomness and spatiotemporal phenomenon. The security and encryption efficiency are also investigated. Results show that the parameter sensitivity can reach the level of $2^{-52}$, the computational precision in practice, when the system size is only three. The parallel encryption is possible even in the one-dimensional coupling structure. Furthermore, the proposed system has excellent resistance against various attacks including the differential-like chosen cipher attack suggested recently. The findings of this paper indicate that the deterministic randomness and its asymptotic model are of great importance for applying in secure communications, and may be extended to other scenarios such as stream cipher and public key cryptography.

## Acknowledgement:

This work was partially supported by NSFC (Grant 60133010, 60102011), NHTP (Grant 2002AA143010, 2003AA143040), and EYTP of Southeast University.

## Reference

[1] L.M. Pecora and T.L. Carroll, Phys. Rev. Lett. 64 (1990) 821.

[2] K.M. Cuomo and A.V. Oppenheim, Phys. Rev. Lett. 71 (1993) 65.

[3] D.G. Vanwiggeren and R. Roy, Science 279 (1998) 1198.

[4] J. Garcia-Ojalvo and R. Roy, Phys. Rev. Lett. 86 (2001) 5204.

[5] S. Wang, J. Kuang, J. Li, Y. Luo, H. Lu and G. Hu, Phys. Rev. E 66 (2002) 065202.

[6] G.N. Tang, S.H. Wang, H.P. Lu and G. Hu, Phys. Lett. A 318 (2003) 388.

[7] H.P. Lu, S.H. Wang, X.W. Li, *et al.*, Chaos 14 (2004) 617.

[8] Y. Zhang, C. Tao, G. Du and J.J Jiang, Phys. Rev. E 71 (2005) 016217.

[9] K.M. Short and A.T. Parker, Phys. Rev. E 58 (1998) 1159.

[10] T. Yang, L.B. Yang and C.M. Yang, Phys. Lett. A 245 (1998) 495.

[11] C. Tao, Y. Zhang, G. Du and J.J Jiang, Phys. Rev. E 69 (2004) 036204.

[12] X.G. Wang, M. Zhan, C.-H. Lai and G. Hu, Chaos 14 (2004) 128.

[13] G.J. Hu, Z.J. Feng and R.L. Meng, IEEE Trans. Circuits Syst-I 50 (2003) 275.

[14] X.W. Li, H.Q. Zhang, Y. Xue and G. Hu, Phys. Rev. E 71 (2005) 016216.

[15] J.A. Gonzalez, B. Lindomar and D. Carvalho, Mod. Phys. Lett. B 11 (1997) 521.

[16] J. Zhou, W. Pei and Z. Y. He, arxiv.nlin.CD/0506026.

[17] Q.Z. Xu, S.B. Dai, W.J. Pei, *et al.*, Neural. Inform. Proc-Lett. & Rev. 3 (2004) 21.

[18] K. Wang, W. Pei and Z. Y. He (unpublished).

[19] J.A. Gonzalez, L.I. Reyes and L.E. Guerrero, Chaos. 11 (2001) 1.

[20] J.A. Gonzalez, L.I. Reyes, J.J Suarez, *et al.*, Phys. Lett. A, 295 (2002) 25.

[21] A. Shamir, B. Tsaban, Inform. Comput., 171 (2001) 350.

Captions of Figures

Fig.1 First return map of the sequence produced by Lissajous map with $a = 5/2$ (a) $f := \sin^2$ (b) $f := \sin$ (c) $f := \cos^2$ (d) $f := \cos$ .

Fig.2 (a) " $x$ distribution- $u$ " bifurcation figures (b) " $y$ distribution- $u$ " bifurcation figures.

Fig.3 (a) Synchronization zone (b) conditional LE vs. coupling strength.

Fig.4 Length of key stream is taken as $L = 2\times 10^3$. (a) cross-correlation of the key stream generated by 24$^{th}$ and 25$^{th}$ lattice of the cryptosystem proposed in Ref. [5]. (b) cross-correlation of the key stream generated by 2$^{nd}$ and 3$^{rd}$ lattice of our proposed cryptosystem.

Fig. 5 The EFA result (a) system proposed in Ref. [5] (b) Our proposed scheme.

Fig. 6 Plaintext recovery. The first 1000 data have been discarded. (a) The plaintext (b) The cipher (c) The recovered plaintext by using $b_1 = 2.5$ (d) The recovered plaintext by using $b_1 = 2.5 - 2^{-51}$ .

Fig.7 The converged state vs. driving. (a) $y_d \in [0,1]$ (b) $y_d \in [0.495, 0.495 + 10^{-10}]$ .

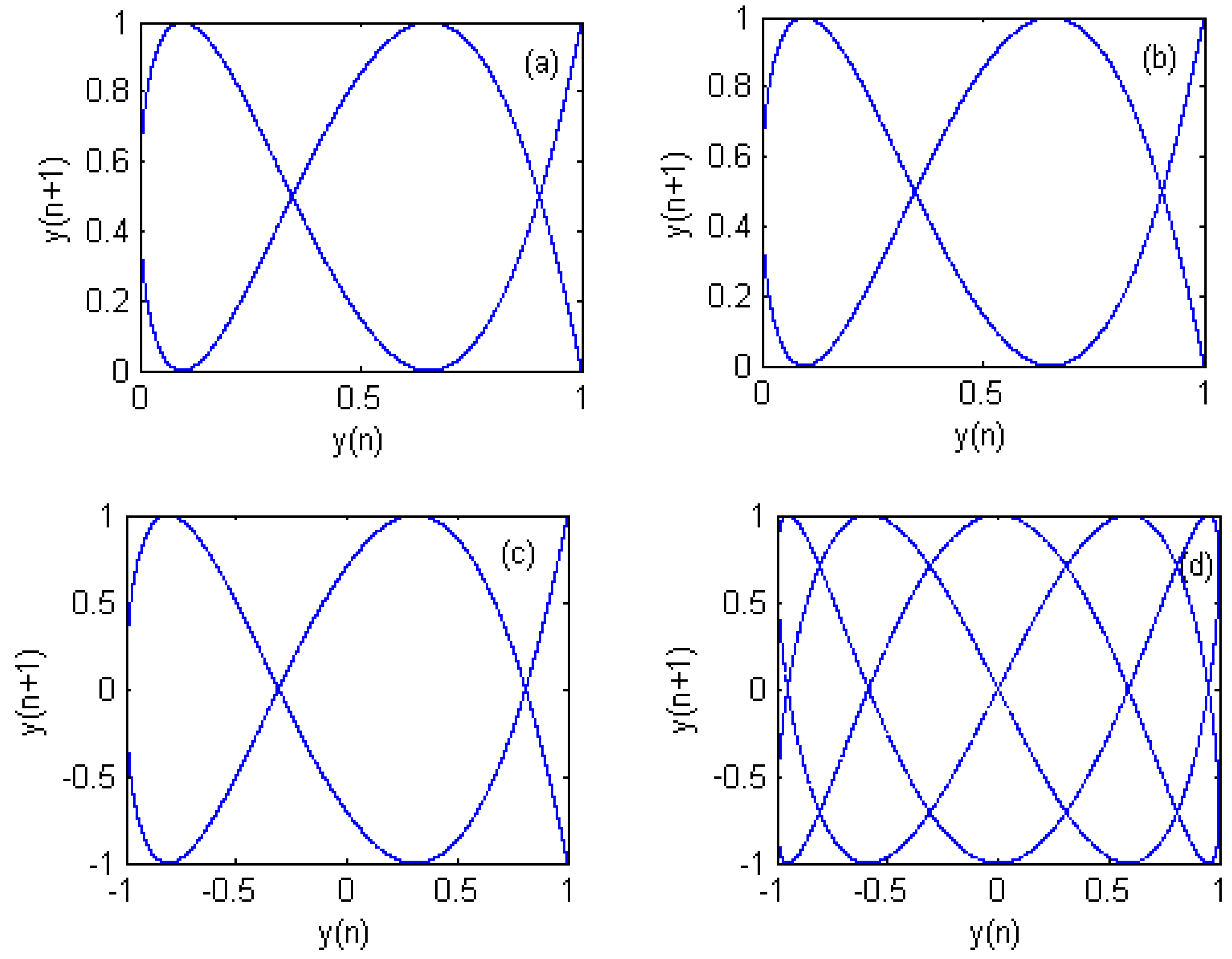

Fig. 1

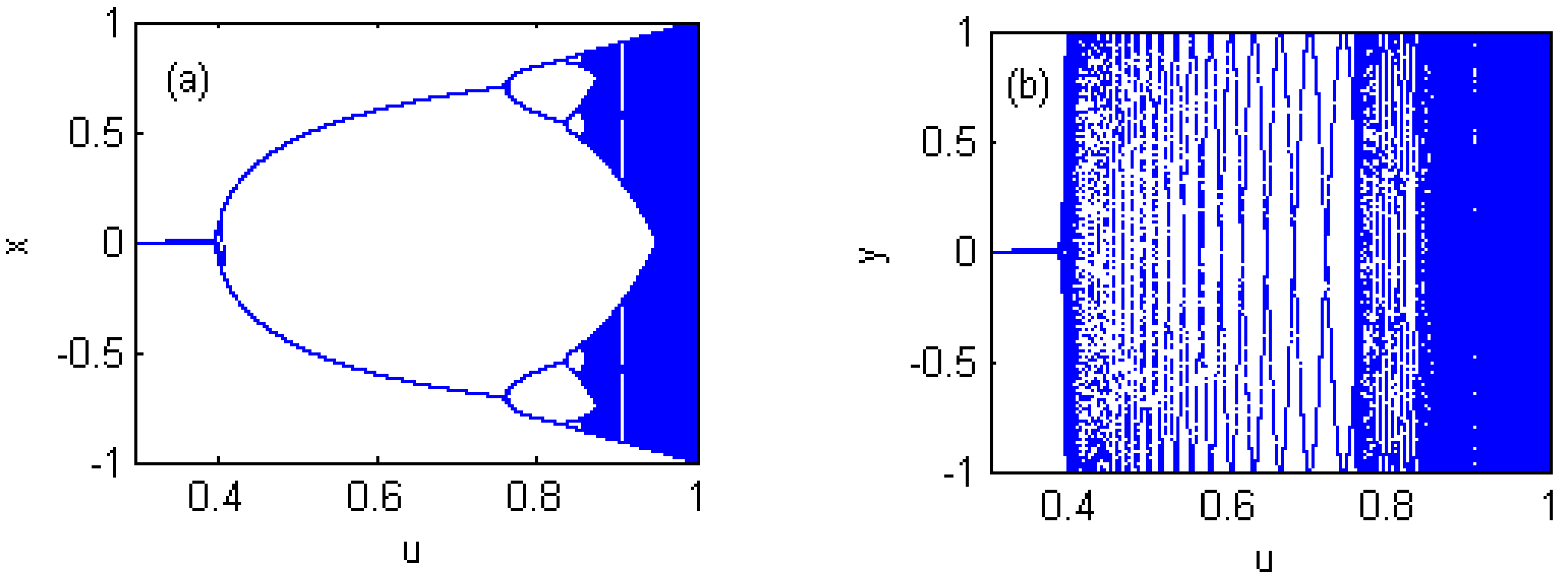

Fig. 2

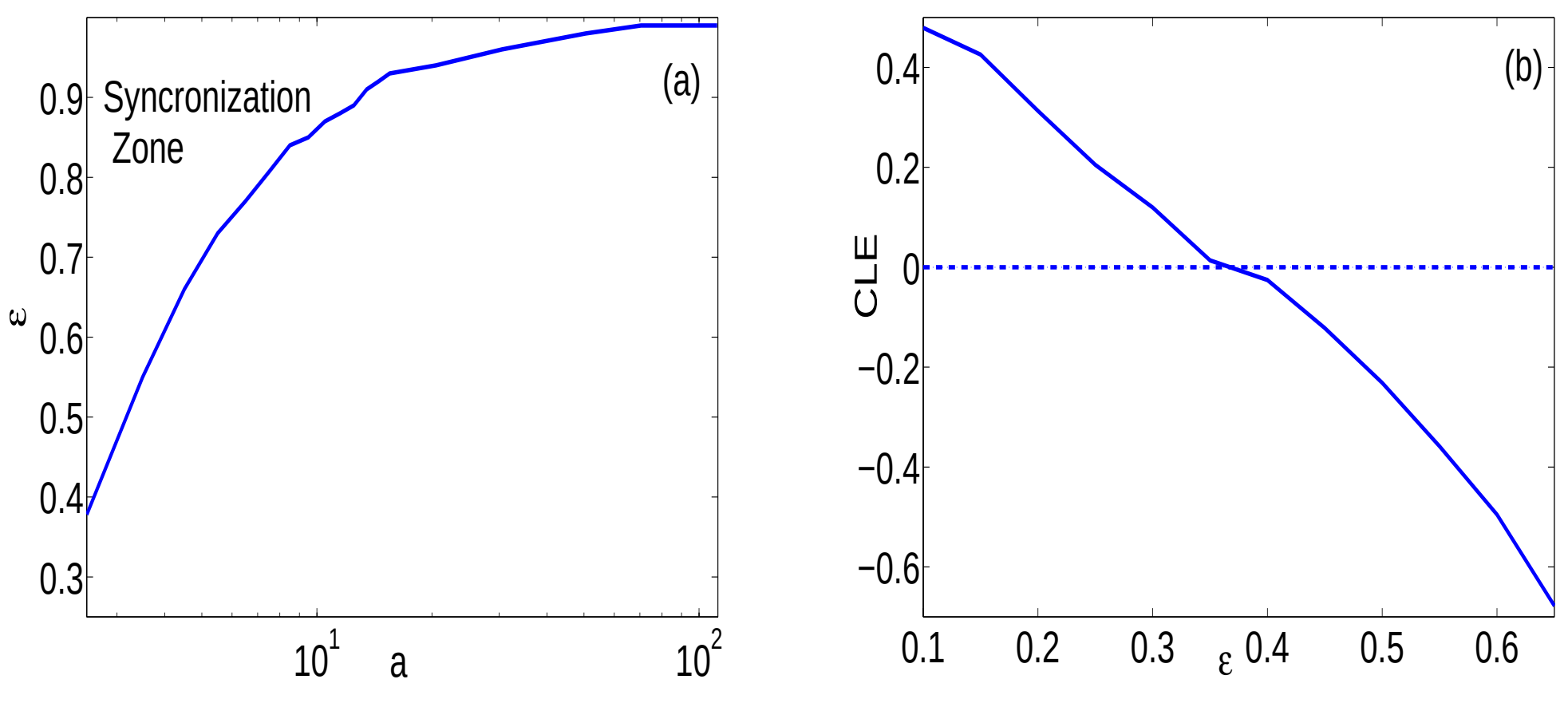

Fig. 3

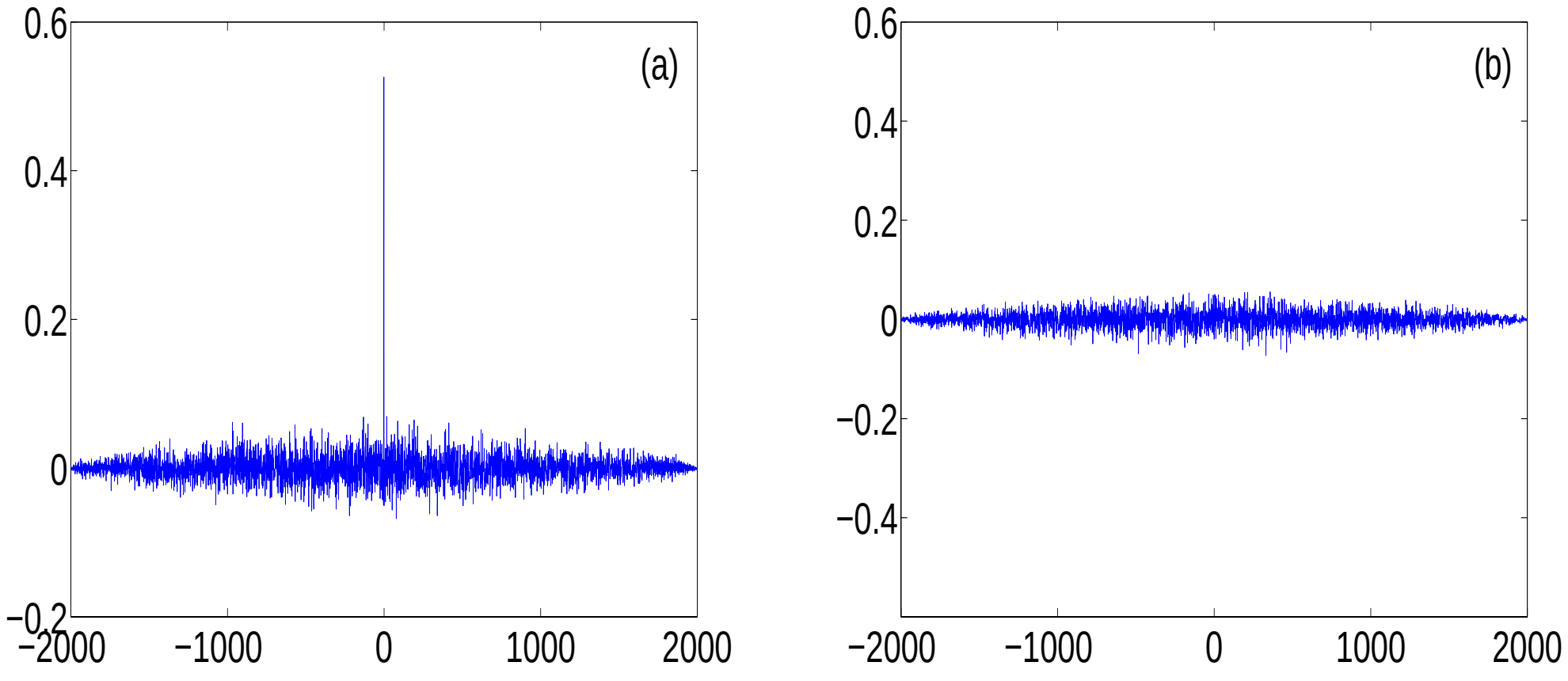

Fig. 4

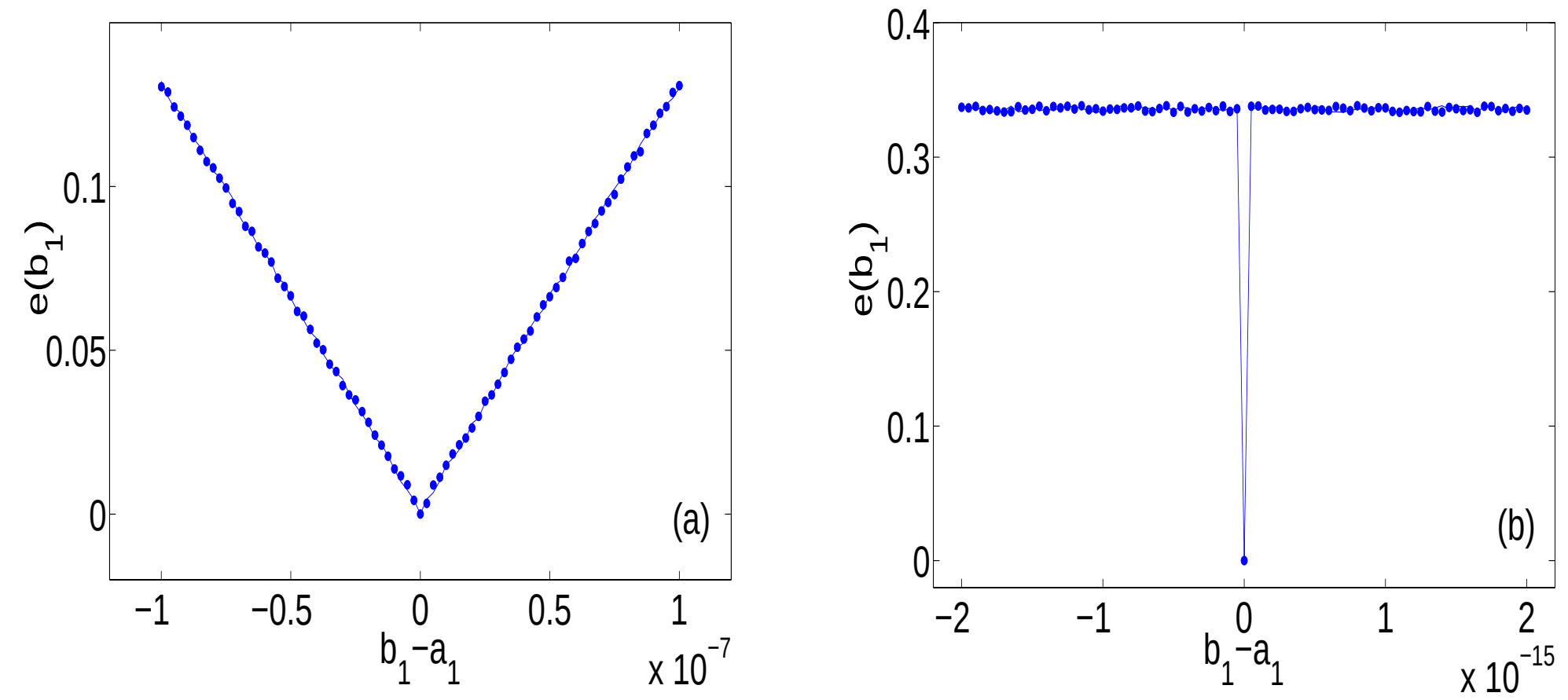

Fig. 5

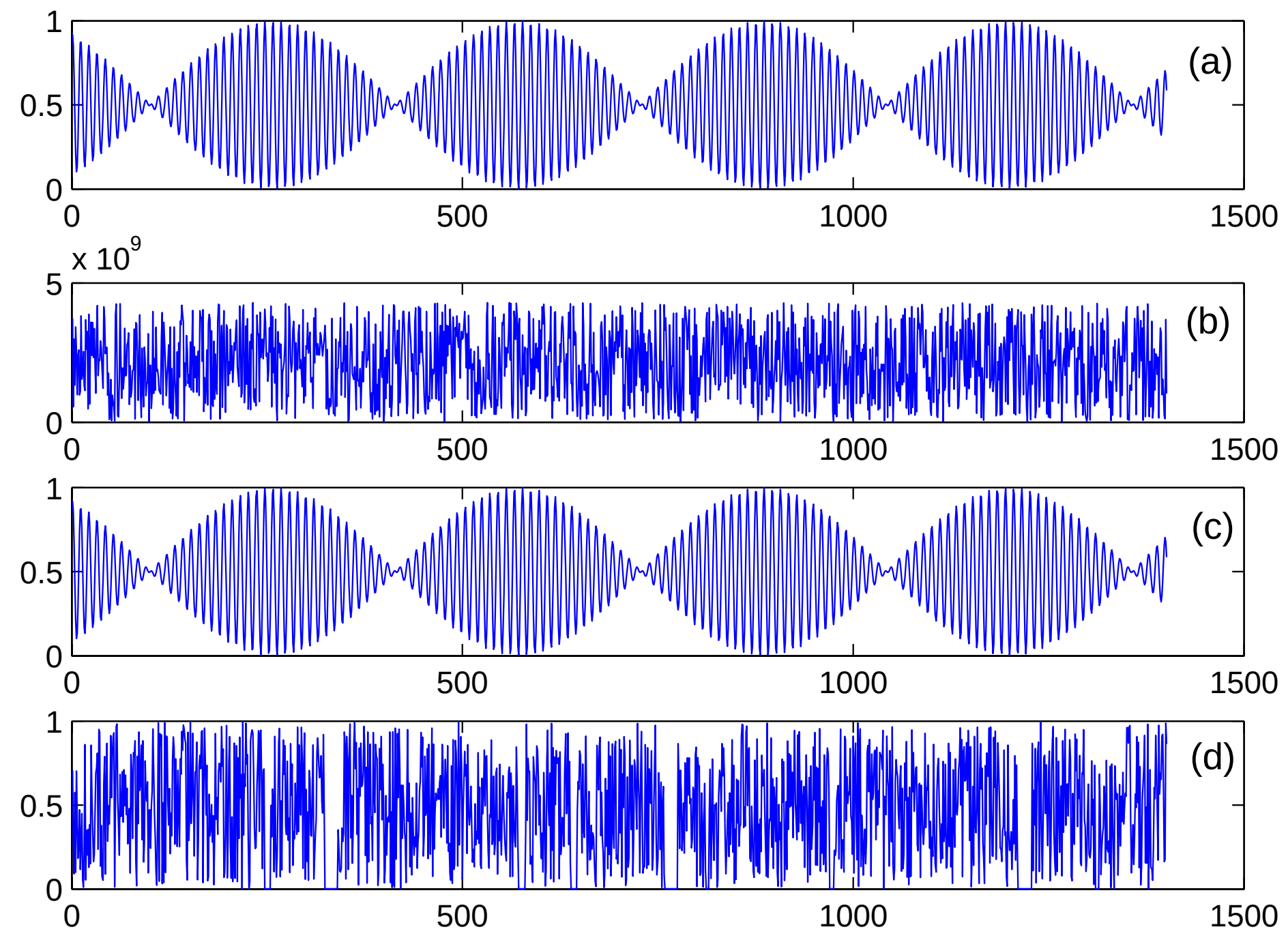

Fig. 6

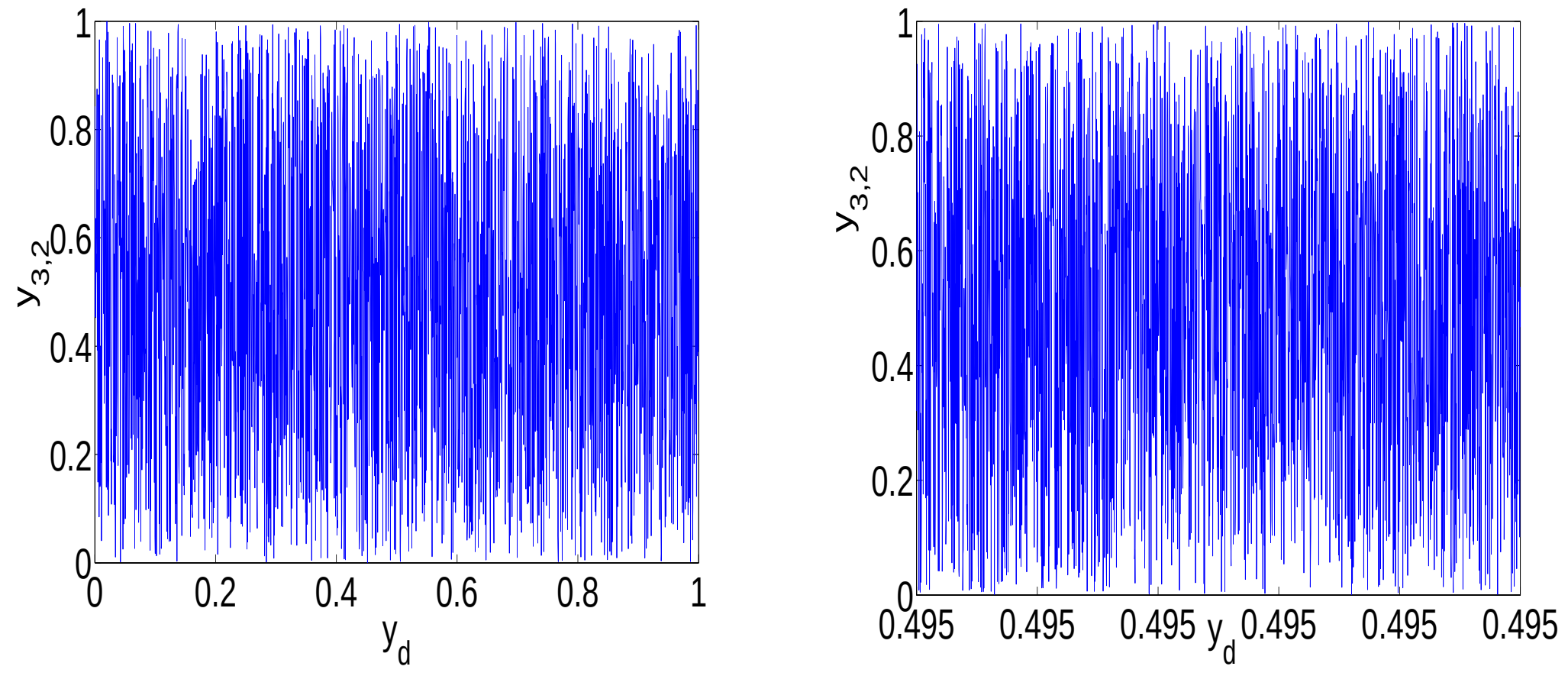

Fig. 7